\documentclass{aa}    
\usepackage{graphicx}
\usepackage{natbib}

\bibpunct{(}{)}{;}{a}{}{,}
\begin{document}

\title{Long-term spectroscopic monitoring of the\\ Luminous Blue
  Variable HD\,160529 \thanks{Based on observations collected at the
  European Southern Observatory at La Silla, Chile (proposals
  69.D-0378, 269.D-5038) }}

\titlerunning{Spectroscopic monitoring of the LBV HD\,160529}

\author{Otmar Stahl\inst{1} \and Thomas G\"ang\inst{2} 
  \and Chris Sterken\inst{3}
  \and Andreas Kaufer\inst{4} 
  \and Thomas Rivinius\inst{1,4} 
  \and Thomas  Szeifert\inst{4} 
  \and Bernhard Wolf\inst{1} 
  }

\offprints{O. Stahl, \email{O.Stahl@lsw.uni-heidelberg.de}}

\institute{Landessternwarte K\"onigstuhl, D-69117 Heidelberg, Germany
  \and L-3 Communications, NASA/GSFC, Greenbelt, MD 20771, USA
  \and Astronomy Group, Vrije Universiteit Brussel, Pleinlan 2, B-1050
  Brussels, Belgium 
  \and European Southern Observatory, D-85748
  Garching, Karl-Schwarzschild-Str. 2, Germany }

\date{Received  / Accepted }

\abstract{We have spectroscopically monitored the galactic Luminous
Blue Variable HD\,160529 and obtained an extensive high-resolution
data set that covers the years 1991 to 2002. During this period, the
star evolved from an extended photometric minimum phase towards a new
visual maximum. In several observing seasons, we covered up to four
months with almost daily spectra. Our spectra typically cover most of
the visual spectral range with a high spectral resolution
($\lambda/\Delta\lambda \approx$ 20\,000 or more). This allows us to
investigate the variability in many lines and on many time scales from
days to years. We find a correlation between the photospheric
He{\sc i} lines and the brightness of the star, both on a time scale
of months and on a time scale of years. The short-term variations are
smaller and do not follow the long-term trend, strongly suggesting
different physical mechanisms. Metal lines also show both short-term
and long-term variations in strength and also a long-term trend in
radial velocity. Most of the line-profile variations can be attributed
to changing strengths of lines. Propagating features in the line
profiles are rarely observed.  We find that the mass-loss rate of
HD\,160529 is almost independent of temperature, i.e.\ visual
brightness.   }

\maketitle

\section{Introduction}

A-type hypergiants and Luminous Blue Variables (LBVs), which often
show spectra similar to A-hypergiants, are the visually brightest
stars in galaxies.  The early A-hypergiant \object{HD\,160529}, which
has long been known to exhibit the spectroscopic signatures of extreme
luminosity \citep{merrill:1943}, has therefore long been considered as
one of the visually brightest stars of the Galaxy, comparable
to the brightest stars in external galaxies.

The variability of HD\,160529 was already established
30\,--\,40\,years ago through studies of its variable light and its
spectrum at different epochs, which indicated changes in its spectral
type.  The spectral variability of HD\,160529 was studied extensively
by \citet{wolf:1974}, who found variations of the Balmer line
profiles, radial velocity variations with an amplitude of about 40\,km
s$^{-1}$\ and line splitting of some metal lines. Intensity
variations of absorption lines of the order of 20\,\% and considerable
photometric variations were also found. The light variability was
studied and described in detail by {\bf  \citet{sterken:1977} and
\citet{sterken:1991}.}

HD\,160529 was classified as a new galactic LBV by
\citet{sterken:1991} due to its brightness decrease of 0.5\,mag from
1983 to 1991 and an apparent spectral type change from A\,9 to B\,8.
Apart from the long-term changes, the authors found pulsation-like
variations in their photometric data with a quasi-period of 57 days
and peak-to-peak amplitudes of 0.1\,mag in $b$ and $y$, while previous
analyses suggested a possible 101.3 day period
\citep{sterken:1981}. More recent photometry has been published by
\citet{manfroid:1991,manfroid:1995} and
\citet{sterken:1993,sterken:1995}.

From comparison with \object{HD\,269662} (=R\,110), a close photometric
and spectroscopic counterpart of HD\,160529 in the Large Magellanic
Cloud, \citet{sterken:1991} derived an absolute visual magnitude of
$M_\mathrm{V} = -8.9$ and a distance of 2.5\,kpc. They estimated the
stellar parameters characterizing the phase of maximum visual
brightness to be $T_\mathrm{eff} \approx 8\,000$\,K, $\log g
\approx\,0.55$, $R_\ast \approx 330$\,R$_\odot$ and $M_\ast \approx
13$\,M$_\odot$.  The derived mass is in good agreement with the mass
determined from evolutionary tracks \citep{lamers:1998}.

\citet{sterken:1991} concluded that HD\,160529 is located near the
lower luminosity limit of the LBV instability strip and possibly in an
evolutionary phase after the Red Supergiant state, i.e.\ in a post-RSG
evolutionary phase.

HD\,160529 is of much lower luminosity than more typical LBVs such as
\object{AG\,Car} \citep{stahl:2001b} and shows much smaller variations
of about 0.5 mag in the visual.  According to
\citet{vangenderen:2001}, it is at the low amplitude limit of the
``strong-active'' S\,Dor variables. The empirical relation between
luminosity and amplitude of LBVs \citep{wolf:1989} is in agreement
with this.  According to the period-luminosity relation of
\citet{stothers:1995} a ``period'' of well above 20 yr would be
expected. This is in rough agreement with the light curve published by
\citet{sterken:1991}.

HD\,160529 can be considered an intermediate case between normal
supergiants and LBVs. The closest counterpart may be the LBV
\object{HD\,6884} (=R\,40) in the SMC \citep{szeifert:1993} or the LBV
HD\,269662 (=R\,110) in the LMC \citep{stahl:1990}. The variability of
HD\,160529 is therefore of interest both in connection with typical
LBVs such as AG\,Car \citep{stahl:2001b} and with normal late B --
early A supergiants \citep{kaufer:1996a,kaufer:1996b}.

Detailed spectroscopic studies of LBVs, which cover both the short and
and long time scales, are still very rare and only available for
AG\,Car \citep{stahl:2001b}. Studies of more objects with different
physical parameters are needed in order to distinguish the general
behaviour of LBVs from peculiarities of single objects. 

\section{Observations}
\subsection{Spectroscopy}

We have observed HD\,160529 with several different echelle
spectrographs and several telescopes from 1991, January to 2002, June.
We have monitored HD\,160529 with high time resolution over a time
span of 2\,--\,4 months in four successive years, i.e.\
1992--1995. In addition, a number of snapshot observations have
been taken to extend the total time base covered.  The epochs of
observations and instruments used are summarized in
Table~\ref{sumtab}.

Most of the observations were obtained with the fiber-linked echelle
spectrographs {\sc Flash} \citep{mandel:1988} and its modified version
{\sc Heros} \citep{stahl:1996}. These instruments were mainly used at
the ESO 50cm telescope, but for a few shorter runs also at the ESO
1.52m and 2.2m telescopes.  In the {\sc Flash} configuration, a
setting of the spectrograph was used that allowed the observation of
the H$\delta$ and He{\sc i}$\lambda$6678 lines with one exposure,
i.e.\ the spectral range covered the wavelength range from about
4\,050\,\AA\ to 6\,780\,\AA\@.  The {\sc Flash} data have been
published on CD-ROM \citep{stahl:1995}.  {\sc Heros} is a modified
version of {\sc Flash}, where a beam splitter is used to divide the
beam after the echelle grating into two channels, each with its own
cross-disperser, camera and CCD detector.  The red channel is
identical to the {\sc Flash} instrument. The two spectral ranges of
the {\sc Heros} configuration cover the range from 3\,450\,\AA\ to
5\,560\,\AA\ in the blue channel and from 5\,820\,\AA\ to 8\,620\,\AA\
in the red channel. The spectral resolution is about 20\,000 for both
{\sc Flash} and {\sc Heros}. The signal-to-noise ratio (S/N) of our
spectra strongly depends on the wavelength and is lowest in the blue
spectral range.  For HD\,160529 we typically used an exposure time of
one to two hours for {\sc Flash} and {\sc Heros} with the ESO 50cm
telescope. In good conditions, a S/N-ratio of at about 100 is reached
in the red spectral range.

Further observations with significantly better S/N and spectral
resolution were carried out with {\sc Feros} \citep{kaufer:2000} at
the ESO 1.52m telescope between 1999 and 2002. The {\sc Feros} spectra
obtained at the ESO 1.52m telescope cover the spectral range from
about 3\,600\,\AA\ to 9\,200\,\AA\ in 39 orders. The spectral resolution
of {\sc Feros} is about 48\,000.  The S/N depends strongly on
wavelength and is highest in the red spectral region. The typical
exposure times were about 10 minutes.

In addition, a few spectra have been obtained with the {\sc Caspec}
echelle spectrograph at the 3.6m ESO telescope and the {\sc Ucles}
echelle spectrograph at the 3.9m AAT telescope. The {\sc Caspec}
spectra cover the spectral range from 3\,800\,\AA\ to 5\,400\,\AA\ with
a spectral resolution of about 20\,000.  With {\sc Ucles},
observations with three grating positions were obtained and the
resulting spectrum covers the total spectral range from 3\,600 to
6\,880\,\AA\@. The spectral resolution is about 50\,000.

For all instruments, a built-in high-temperature incandescent lamp and
a Th-Ar lamp were used to obtain flat-field and wavelength-calibration
exposures, respectively.

In addition, we obtained on Sep. 24, 2001, low resolution spectra
(3\,\AA/pix in the wavelength range from 4\,000 to 8\,000\,\AA) with
the {\sc Dfosc} focal reducer spectrograph at the Danish 1.54m
telescope at ESO, La Silla. In this case, a He-Ne lamp was used for
the wavelength calibration. The {\sc Dfosc} data are of much lower
resolution than the echelle data and allow to study the equivalent
width variations only. Since the spectra were obtained in
non-photometric conditions, they can also not be flux-calibrated.

The {\sc Ucles} spectra were reduced with IRAF\@. All other spectra were
reduced with ESO-Midas. The {\sc Caspec} spectra were reduced with the
standard echelle package of ESO-Midas \citep{ponz:1983}. For the {\sc
Flash} and {\sc Heros} spectra a modified version of this package
\citep{stahl:1993} was used.  The {\sc Feros} package
\citep{stahl:1999}, also running within ESO-Midas, was used for the
reduction of the {\sc Feros} spectra. The {\sc Dfosc} spectra have
been reduced with the long-slit package of ESO-Midas.

\begin{table}
  \begin{center}
    \caption[]{Summary of the spectroscopic observations of HD\,160529.
    }
    \begin{tabular}{llrll}
      \hline
      {\bf Instr.} & {\bf Telescope} & {\bf Sp.} & {\bf year} \\        
      \hline
      {\sc Caspec}    &  ESO 3.6m         & 2       & 1991.08         \\  
      {\sc Flash}     &  ESO 50cm         & 62      & 1992.47-1992.75 \\  
      {\sc Flash}     &  ESO 50cm, 2.2m   & 72      & 1993.09-1994.41 \\  
      {\sc Flash}     &  ESO 50cm         & 55      & 1994.14-1994.49 \\  
      {\sc Heros}     &  ESO 50cm         & 23b/27r & 1995.22-1995.42 \\  
      {\sc Ucles}     &  AAT 3.9m         & 1       & 1995.76         \\  
      {\sc Heros}     &  ESO 1.52m        & 6       & 1997.30-1995.32 \\  
      {\sc Feros}     &  ESO 1.52m        & 2       & 1999.53-1999.56 \\  
      {\sc Dfosc}     &  Danish 1.54m     & 3       & 2001.74         \\  
      {\sc Feros}     &  ESO 1.52m        & 2       & 2002.49 52455   \\  
      \hline
    \end{tabular}
    \label{sumtab}
  \end{center}
\end{table}

In the years 1992--1995, the star was observed at least once per week
and in some seasons up to once per night. In these runs, the sampling
is sufficiently dense to cover variations with time scales of a few
days.

\subsection{Photometry}

\begin{figure}
  \resizebox{8.8cm}{!}{\includegraphics[angle=-90]{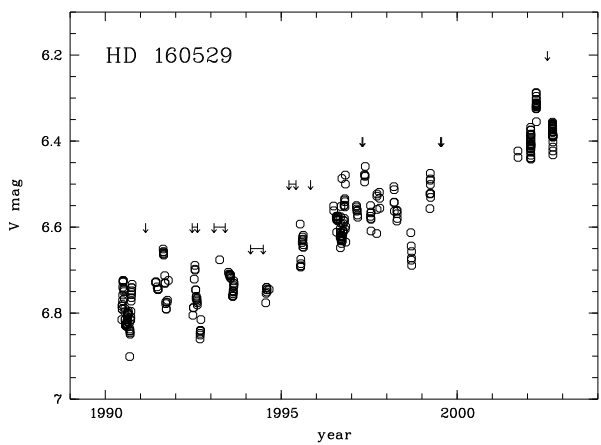}}
  \caption[]{Results of the photometry of HD\,160529 in the
  Str\"omgren $y$ band. {\bf The x-axis is in years.}  The epochs of
  spectroscopic observations are indicated by arrows, longer observing
  runs are indicated by double arrows. During our observations, the
  star evolved from photometric minimum to a new maximum.}
  \label{photo_fig}
\end{figure}

Differential photometry was carried out in the framework of the LTPV
project \citep{sterken:1983} using the ESO 50cm telescope (PMT
photometry), the Dutch 90 cm telescope (CCD) and the Danish 1.54m
telescope (CCD) at ESO, La Silla. Zero point corrections were applied
but no proper standardization was attempted. The results based on the
measurements in the Str\"omgren $y$ band are displayed in
Fig.~\ref{photo_fig}.

\section{Variability}
\subsection{Changes in spectral type}

The large photometric variations of LBVs are typically accompanied by
changes of the spectral type, which can change from Of/WN- to
A-hypergiant \citep{stahl:2001}. For HD\,160529, spectral types
between B8Ia and A9Ia have been published in the literature. However,
the spectrum of HD\,160529 is peculiar and, since the spectral type
depends on the exact classification criteria used, it is not clear if
the differences in the published spectroscopic classifications are due
to real variations. Most authors find a spectral type of A2--A3
\citep{wolf:1974}, only two authors find a strongly deviating type:
\citet{houk:1982} finds A9Ia and \citet{sterken:1991} infer spectral
type variations from B8 to A9.

It should be noted, that the spectral types published by
\citet{sterken:1991} are not based on spectroscopic criteria, but they
are {\em inferred\/} from two arguments:

\begin{itemize}
\item The observed color variations variations indicate temperature
  changes from 8\,000 -- 10\,000 K. This assumes that the theoretical
  colors apply to the extreme supergiant HD\,160529.
\item The visual amplitude of 0.5 mag is due to a change in the
  bolometric correction at constant bolometric luminosity. This
  assumes that the bolometric corrections
  \citep{schmidt-kaler:1982} for supergiants apply to HD\,160529.
\end{itemize}

The classification (A9Ia) of \citet{houk:1982} is therefore the only
spectroscopic classification strongly deviating from the A2--A3 range.
The remark of \citet{houk:1982}: ``H lines seem even weaker than in
standards; partly filled in?'' seems to indicate that this
classification is partly based on the weakness of the Balmer lines,
which are indeed influenced by emission.
  
\begin{figure}
  \resizebox{8.8cm}{!}{\includegraphics[angle=0]{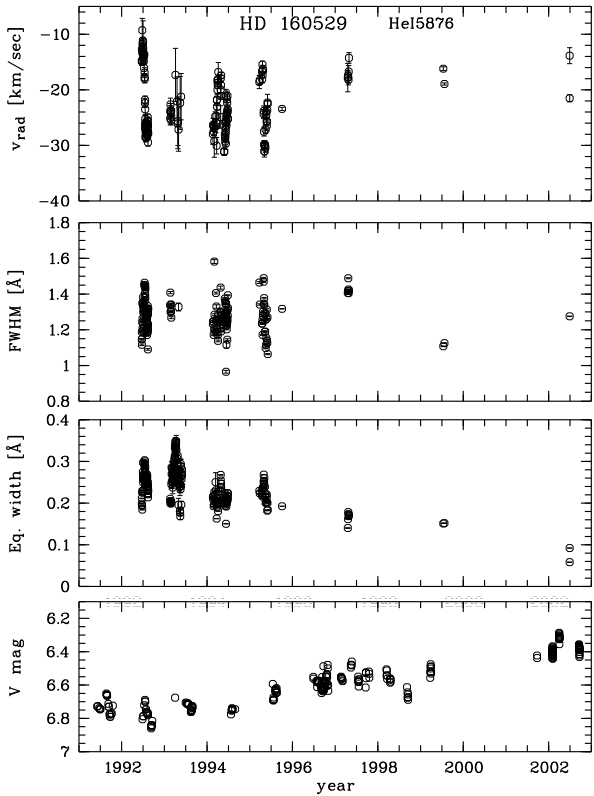}}
  \caption[]{Radial velocity, FWHM and equivalent width of He{\sc
    i}$\lambda$5876 as a function of time. The equivalent width of
    this line has been measured for other supergiants to 0.68\,\AA\ for
    B8Ia, 0.43\,\AA\ for B9Ia, 0.24\,\AA\ for A0Ia and 0.10\,\AA\ for
    A2Ia. By this criterion, the star is coolest in 2002.}
  \label{HeI5876_fig}
\end{figure}

We therefore re-evaluated the evidence for strong changes in the
spectral type of HD\,160529. From our spectra we infer only minor
changes in spectral type for the spectra obtained until 1999.  For
this period, we determined the spectral type of HD\,160529 by
comparing with the supergiant stars $\beta$\,Ori (B8Ia), HD\,96919
(B9Ia), HD\,92207 (A0Ia) and HD\,100262 (A2Ia) from the sample of
\citet{kaufer:1997b}. The match of most lines is reasonably good with
an approximate spectral type of A2Ia. The spectral type as determined
from the strength of the metal lines is certainly always later than A0
and earlier than A5Ia.  Since most of the metal lines are forming
partly in the expanding atmosphere of HD\,160529, it is not clear how
this spectral type relates to the star's effective temperature. This
spectral type of A2Ia is in disagreement with the strength of He{\sc
i}$\lambda$5876, which would indicate a spectral type of B9Ia --
A2Ia. The given range reflects the strong variations of this line
(cf.~Fig.~\ref{HeI5876_fig}).

We find that the spectrum before 1999 was significantly different from
the maximum phase spectra of \citet{wolf:1974}. The spectral type was
earlier (hotter) than in maximum phase. In addition, the emission
lines are stronger in minimum phase. The temperature change is
inferred mainly from the weakening of Ti{\sc ii} and Cr{\sc ii} lines
in photometric minimum. However, since these lines form also in the
stellar wind, the spectral type changes are unreliable indicators for
the effective temperature. The He{\sc i}$\lambda$5876 line may be a
better temperature indicator than the metal lines, although the
variable profile of He{\sc i}$\lambda$5876 indicates that this line is
also not of purely photospheric origin.

\begin{figure}
  \resizebox{8.8cm}{!}{\includegraphics[angle=-90]{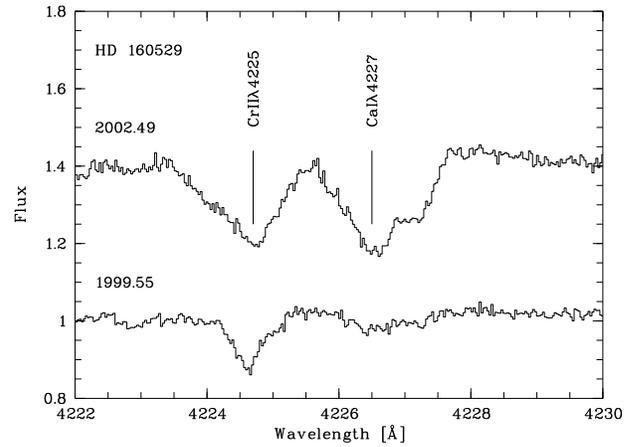}}
  \caption[]{Variation of the Ca{\sc i}$\lambda$4227 line.  The
    strength of this line strongly increased in 2002.}
  \label{CaI_fig}
\end{figure}

Nevertheless, part of the spectral changes can certainly be attributed
to changes in temperature. The strength of most metal lines strongly
increased in 2002. In particular the Ca{\sc i}$\lambda$4227 line,
which had equivalent widths around 50\,m\AA\ in earlier years,
increased to about 270\,m\AA\ in 2002 (cf.~Fig.\ref{CaI_fig}.
This line is a good temperature
indicator for A stars. The strength of this line is slightly smaller
than in the LBV \object{HD\,269662} (=R\,110) in 1989, which at that
time was classified as about F0Ia \citep{stahl:1990}. The evidence is
strong that the star in 2002 was considerably cooler than in the years
before. The spectral type was earlier than F0, but much later than A2
and probably late A, confirming the spectral type variations reported
earlier.

\subsection{Variations of line strengths and radial velocity}

Since HD\,160529 developed a new S\,Dor phase during our observing
campaign, strong spectroscopic changes are to be expected.

\begin{figure}
  \resizebox{8.8cm}{!}{\includegraphics[angle=-90]{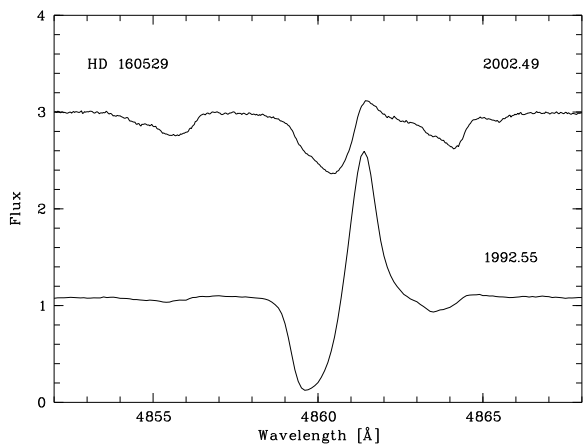}}
  \caption[]{The H$\beta$ line and two Cr{\sc ii} lines in 1992 
    (bottom) and 2002 (top). Note the much weaker emission and
    absorption components of H$\beta$ and the stronger metal
    absorption lines in 2002 as compared with 1992. }
  \label{var_Hbeta}
\end{figure}

The general appearance of the spectrum of HD\,160529 changes very
little within one observing season. Much stronger variations were
observed during the whole campaign from 1992 to 2002. The lines of
Cr{\sc ii} and Ti{\sc ii} are very temperature-sensitive in the
temperature range of HD\,160529 (i.e.\ about 8\,000 to 12\,000\,K) and
temperature changes should be reflected in the {\em strength\/} of
these lines.  Indeed, these lines exhibit particularly strong
variations during our observing campaign.  However, the line {\em
profiles\/} are also highly variable, and {\bf is is therefore not
clear that the changes in line strength are due to temperature changes
only. Especially the short-term changes could be significantly
affected by stellar wind effects (cf.~e.g.\ Fig~\ref{CrII_2_fig}),
while the long-term changes, where these effects are averaged out, are
more likely to be dominated by temperature variations.}
Fig.~\ref{var_Hbeta} shows an example of the spectral changes in
Cr{\sc ii} lines. The Cr{\sc ii} are also sharper in 2002, which is
also visible when comparing with Fig.~7 of \citet{sterken:1991}: they
show a spectrum obtained in 1990, where the Cr{\sc ii} lines have a
strong blue-shifted component.  Compare with Fig.~4 of
\citet{wolf:1974}, where the Cr{\sc ii} lines are always strong. Also
by comparing our equivalent width measurements with Table~6 of
\citet{wolf:1974}, we find much lower values for Cr{\sc ii} and Ti{\sc
ii} in most years. Only in 1999 and 2002, our values approach the
values of \citet{wolf:1974}. The weakening of Ti{\sc ii} during
minimum has also been reported by \citet{sterken:1991}.

{\bf Fig.~\ref{var_Hbeta} also shows strong variations of the H$\beta$
line, both in strength and line profile, between 1992 and 2002. The
line strength variations are probably caused by a density decrease in
the stellar wind, caused by a radius increase and/or a decrease of the
stellar radius. The line profile shows that, while the terminal
velocity is unchanged, the minimum of the absorption moved to smaller
expansion velocities. This could be due to decreased optical depth in
the H$\beta$ line or a decrease of the stellar wind velocity in the
inner part of the wind.  }

\begin{figure}
  \resizebox{8.8cm}{!}{\includegraphics[angle=0]{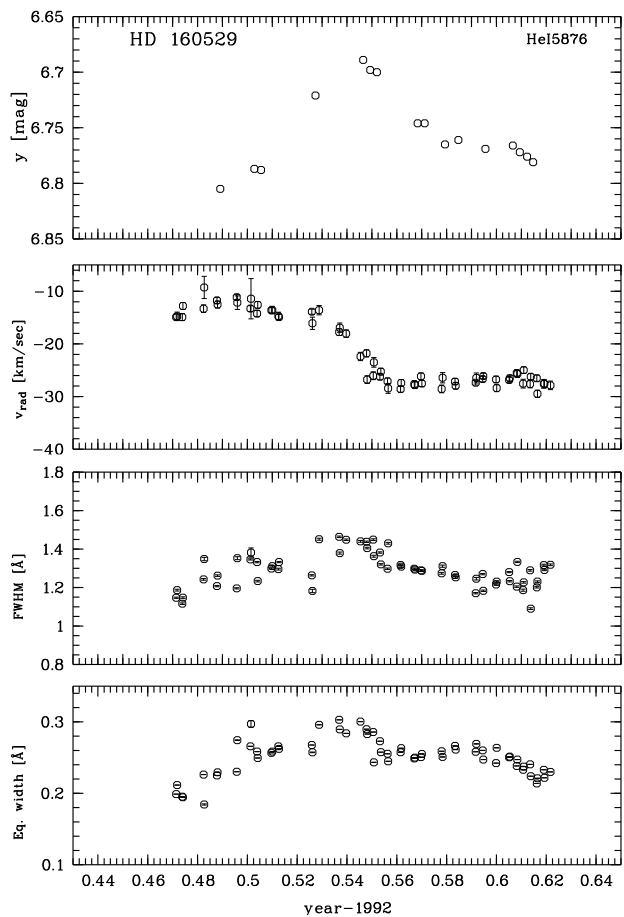}}
  \caption[]{Photometric and spectroscopic variations of the He{\sc
    i}$\lambda$5876 line in 1992. Note the fast changes of almost 15
    km s$^{-1}$ within only about 10 days and the relative constancy
    before and after this change. The equivalent-width curves seems to
    follow the light curve.}
  \label{HeI5876_2_fig}
\end{figure}

The He{\sc i}$\lambda$5876 line is relatively symmetric and thus
appears well-suited to study the photospheric variations of
HD\,160529.  The radial velocity of this line is highly variable, but
also the FWHM and equivalent width of the line change. We fitted
Gaussian profiles to all spectra and so obtained not only the radial
velocity but also the residual intensity and line width as a function
of time.  The results are shown in Figs.~\ref{HeI5876_fig}
and~\ref{HeI5876_2_fig}. It is obvious from this figure that the line
is not only variable in radial velocity but also in line width and
intensity.  Radial velocity and equivalent width are not correlated.
The equivalent width and radial velocity are highly variable within
each observing season, i.e.\ on time scales of a few months. In
addition, the average equivalent width seems to decrease from 1995 to
2002. This probably indicates a lower temperature in these years which
is supported by e.g.\ Ti{\sc ii} line measurements. The H$\alpha$
strength also changed strongly in 1999--2002. The equivalent width was
varying between about 14 and 17\,\AA\ from 1992 to 1997. In 1999 it
was down to 10\,\AA\ and further decreased to 7.5\,\AA\ in 2001 and to
6.5\,\AA\ in 2002.

The relation of spectroscopic behaviour and light curve on short time
scales can only be studied from the spectra obtained in 1992, when we
have sufficient overlap in the spectroscopic and photometric data set
(cf.~Fig.~\ref{HeI5876_2_fig}). The sudden decrease of radial velocity
seems to coincide with a peak of the visual brightness. The equivalent
width of He{\sc i}$\lambda$5876 and also the FWHM have a peak at the
same time. Note that this correlation of line strength with visual
brightness on short time scales contrasts with the pronounced {\em
anti-correlation\/} on long time scales (cf.~Fig.~\ref{HeI5876_fig}).
{\bf These short-term variations of the He{\sc i}$\lambda$5876 line
could be due to pulsational variations or to stellar-wind effects. If
they are due to pulsation, the variations of the equivalent width
imply an increase of the temperature with brightness. If we
interpret the variations as changes of the stellar wind, the
observations would suggest a strengthening of the wind with
brightness, with a phase-delay of the radial velocity.}

Other lines which mainly originate in the photosphere are e.g.\ Mg{\sc
  ii}$\lambda$4481 and the Si{\sc ii}$\lambda\lambda$6347, 6371 lines.

If variations in the strength of the Ti{\sc ii}, Cr{\sc ii} and He{\sc
i} lines are due to photospheric temperature change of the star, we
would expect the changes in He{\sc i} and the metal lines to be
anti-correlated. Indeed, there is a weak evidence for this behaviour
in some years, but the scatter is very large within individual years.
In the long-term trend, the anti-correlation is clearly visible.

\begin{figure}
  \resizebox{8.8cm}{!}{\includegraphics[angle=0]{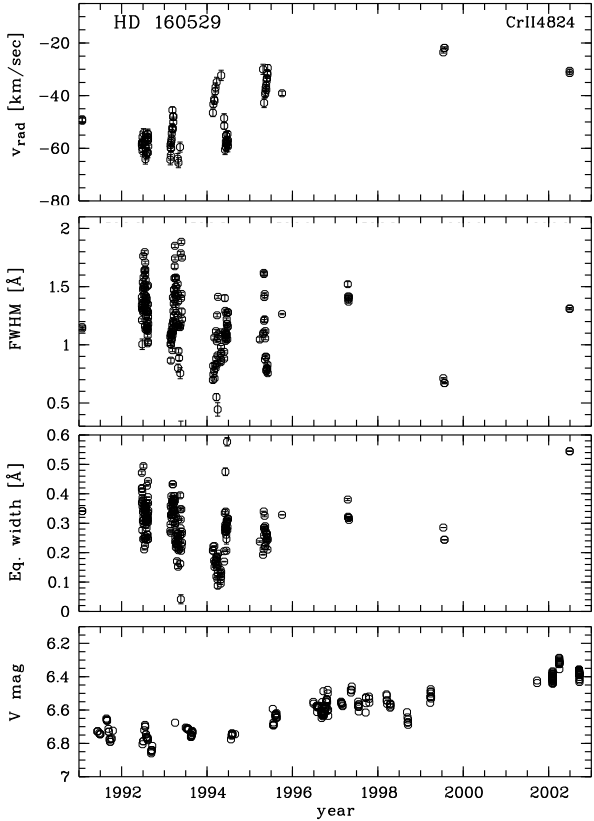}}
  \caption[]{Radial velocity, FWHM and equivalent width of Cr{\sc
    ii}$\lambda$4824 as a function of time. The parameters have been
    determined by Gaussian fits to the line profiles, although the
    line profiles often are not Gaussian. Note the
    long-term trend in the radial velocity and the extremely strong
    changes in equivalent width on a time scale of weeks. The average
    equivalent width of about 0.3\,\AA\ indicates a spectral type of
    about A2Ia, 0.1\,\AA\ indicates A0Ia.}
  \label{CrII_fig}
\end{figure}

\begin{figure}
  \resizebox{8.8cm}{!}{\includegraphics[angle=0]{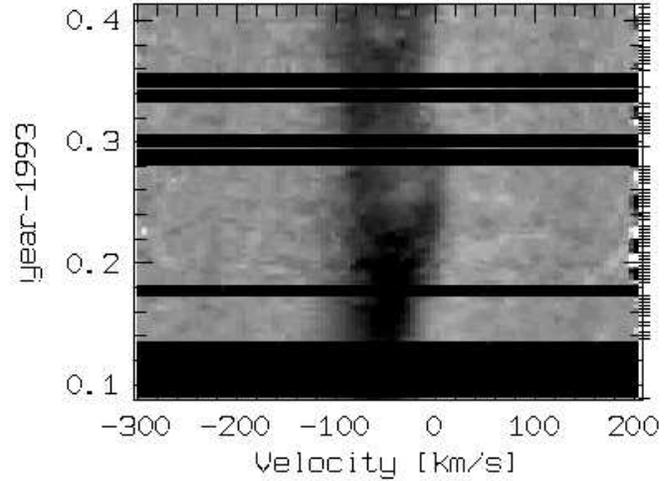}}
  \caption[]{Profile variations of Cr{\sc ii}$\lambda$4824 as a function 
    of time in 1993. Note the strong change in line width and the
    occasional line-splitting. }
  \label{CrII_2_fig}
\end{figure}

This is probably due to the fact that the metal lines form mostly in
the stellar wind (see Figs.~\ref{CrII_fig} and~\ref{CrII_2_fig}). Only
at the end of the observing run, the lines appear close to the
systemic velocity and with a line width that is compatible with $v
\sin i$. In other years, the radial velocity, line width and line
profile variations indicate a stellar wind origin for these lines.
Line splitting is quite common and has already been observed by
\citet{wolf:1974}. In particular in the first years of the
observations, the radial velocities are more negative than observed by
\citet{wolf:1974} in maximum visual brightness.

\subsection{Time series analysis}

\citet{lamers:1998} have previously analyzed the photometric data of
HD\,160529. For two seasons, they derived a period of 45 and 55 days,
respectively. They explained the variations by non-radial $g$-mode
pulsations of low order $l$ and the period change by mode-switching.

We examined the photometric data and radial velocity measurements for
periodicities with the ESO-Midas TSA package using the method of
\citet{scargle:1982}.  We limited the measurements to seasons with a
significant number of measurements. For the photometry, these were the
years 1991--1993 and for the spectroscopy the years 1992--1995.  We
restricted the search to periods below 120 days, since this is the
approximate length of the observing runs. For the $y$-photometry we
obtain the strongest peaks in the power spectrum at about 108, 83 and
68 days and for the radial velocity of He{\sc i} at 93 and 67
days. The equivalent width of this line shows two peaks at similar
periods at 92 and 67 days and in addition two peaks at 126 and 82
days.  We checked that the results do not depend critically on the
choice of the data set. If data sets from individual years are removed
or added, the detected periods change, but we always find peaks at
similar periods, {\bf i.e.\ within a few days.  This could indicate
that the variations are not strictly periodic or not coherent over
longer time scales. However, the small changes of the position of the
peaks of the power spectrum are more likely due to the irregular
sampling of our data. }

It should also be mentioned that the radial velocity changes are
definitely not sinusoidal. Although the time series analysis gives
time scales of around 100 days, scales for the changes are of the
order of 10 days. The time series analysis measures the time scale on
which such sudden changes of the radial velocity repeat.  Examples of
the observed variations are shown in Figs.~\ref{HeI5876_fig}
and~\ref{HeI5876_2_fig}.  Similar events have been observed in other
years as well.

The similarity between the time scales of the photometric and
spectroscopic variations suggests a physical connection.  {\bf In 1992,
when} we obtained photometry and spectroscopy almost simultaneously, a
peak in the light curve was observed to coincide with the a rapid
change in radial velocity (Fig.~\ref{HeI5876_2_fig}). The equivalent
width of He{\sc i}$\lambda$5876 also has a peak at the same time.  In
fact, the equivalent width seems to follow quite closely the visual
light curve. Unfortunately, our observations do only allow to observe
these relation in the 1992 observing seasons. Therefore, we can only
conclude that the rapid changes in radial velocity {\em may} \/be
related to the photometric variability.

\subsection{Line profile variations}

Many lines in the spectrum of HD\,160529 show peculiar line profile
variations. While this was already known from the work of
\citet{wolf:1974}, the time sampling of the spectra of these authors
was not adequate to follow the changes in line profiles in detail.

We searched in particular for propagating features in the line
profiles, in order to study the stellar wind of HD\,160529.  Most of
the lines are dominated by absorption and the observed {\bf
variability seems mainly due} to additional absorption.  In order to
increase the contrast, we therefore examined the residuals of the
profiles with respect to a ``maximum flux'' profile, which was
constructed from the time series by computing the highest profile for
each wavelength bin in each observing season. {\bf This procedure is
very sensitive to noise. Therefore, a running median was applied along
the $t$-axis before computing the maximum flux.  The subtraction of a
``maximum flux'' profile is more appropriate than subtracting a mean
profile if the variations are due to additional absorption on top of a
stable profile. The subtraction of a mean profile always introduces
emission-like features in the residuals. }

\begin{figure}
  \resizebox{8.8cm}{!}{\includegraphics[angle=0]{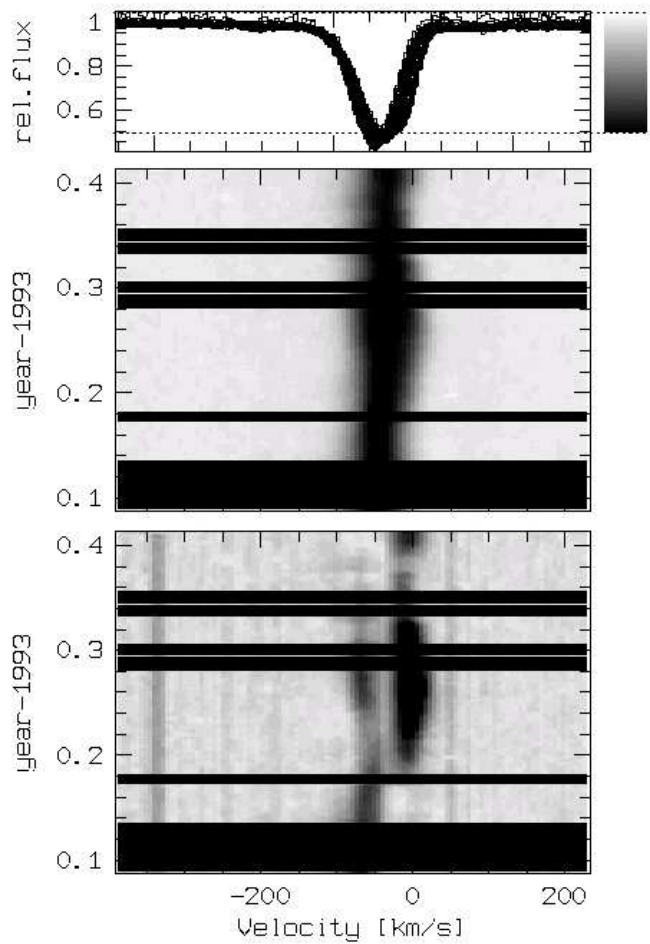}}
  \caption[]{Profile of the Si{\sc ii}$\lambda$6347 line in 1993.
    A ``maximum flux'' profile has been subtracted in the lower panel
    to enhance the contrast. }
  \label{SiII6347_fig}
\end{figure}

A number of strong absorption lines {\bf do not show any emission
component or a strong wind absorption and, therefore, superficially
look like photospheric lines.}  An example of such metal lines are the
Si{\sc ii} lines at 6347, 6371\,\AA\@. Profile variations as observed
in 1993 are shown in Fig.~\ref{SiII6347_fig}.

Most of the stronger metallic lines, in particular Fe{\sc ii} lines,
show pronounced P\,Cyg profiles. These lines show peculiar variations
of the line profiles. An example of the variations is shown in
Fig.~\ref{FeII6248_fig}. Also in these profiles, most of the
variability can be ascribed to additional absorption of variable
strength, but with little change in radial velocity.  The variability
is strongest at {\bf very low velocities around 0 km s$^{-1}$. This
component appears directly on top of the emission component. Therefore
the P~Cyg profile changes to a splitted absorption profile (see
Fig.~\ref{FeII6248_fig}, top) and later back to a P~Cyg
profile. Additional absorption at blue-shifted velocities of about $-50$
km s$^{-1}$ is also observed, but the features at low and high
expansion velocities do not seem to be connected. This second,
blue-shifted component seems to move to higher expansion velocities,
but the changes are quite small. In some cases, the absorption
features even seem to move towards longer wavelengths.  Features which
are clearly moving to higher expansion velocities, i.e.\ towards
shorter wavelengths, are not often observed. Therefore, while we have
some evidence for structures which originate in an accelerating
stellar wind, the strongest variations occur close to the systemic
velocity. A cool photospheric absorption layer may be responsible for
these variations.}

\begin{figure}
  \resizebox{8.8cm}{!}{\includegraphics[angle=0]{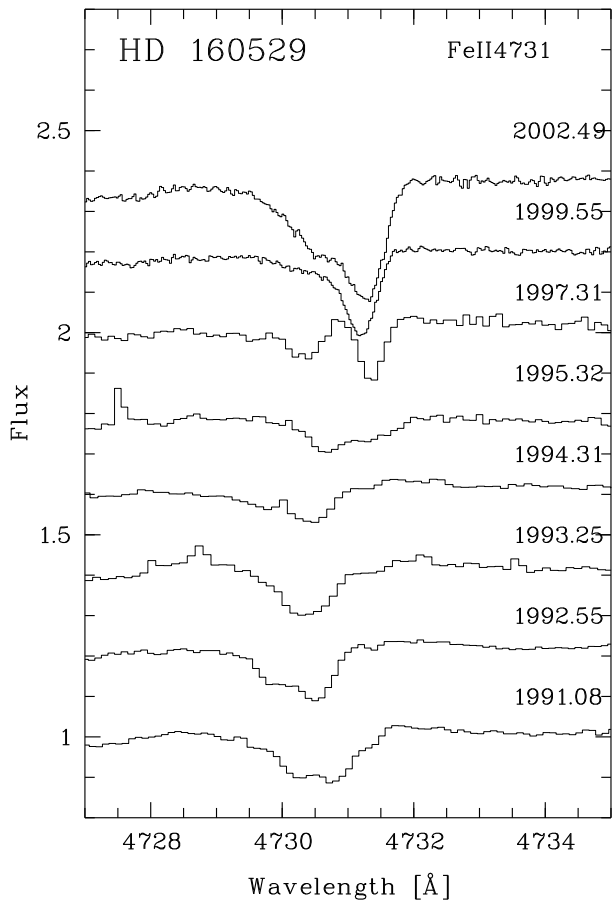}}
  \caption[]{Profile of the Fe{\sc ii}$\lambda$4731 line in various
    observing seasons from 1991 to 2002. The labels of the spectra
    denote {\bf the date in years.} Note the strong low-velocity
    feature in 1997 and the strength and narrowness of the feature in
    1999. In 2002, the strength of the line further increases and a
    strong blue wing develops.}
  \label{FeII4731_fig}
\end{figure}

\begin{figure}
  \resizebox{8.8cm}{!}{\includegraphics[angle=0]{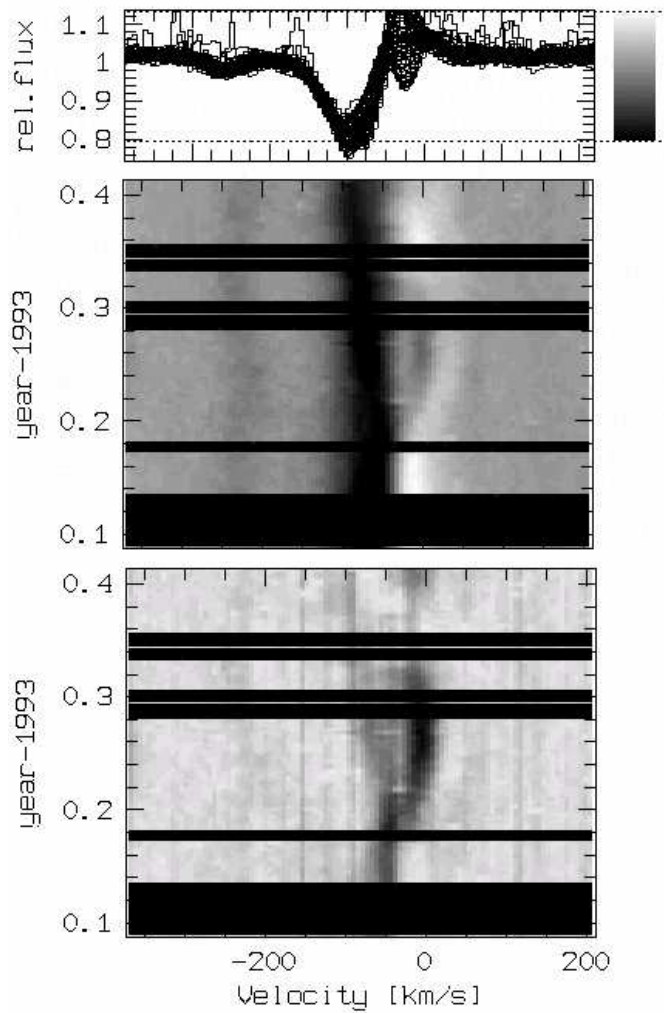}}
  \caption[]{Profile of the Fe{\sc ii}$\lambda$6248 line in 1993. 
    A ``maximum flux'' profile has been subtracted in the lower panel
    to enhance the contrast. Note the striking similarity with
    Fig.~\ref{SiII6347_fig}. }
  \label{FeII6248_fig}
\end{figure}

The low-velocity feature is particularly strong in 1997, where it is
clearly seen as a separate absorption feature. As an example we show
in Fig.~\ref{FeII4731_fig} the profile of the Fe{\sc ii}$\lambda$4731
line in the various observing seasons. In this figure it can be seen
that the line was clearly blue-shifted from 1991 to 1994 and later
moved to longer wavelengths.  The low-velocity feature could be
related to the slightly red-shifted absorption features seen sometimes
in LBVs \citep{wolf:1990,stahl:2001}.

A comparison of the line profile variations of lines which have a
photospheric appearance (e.g.\ Si{\sc ii}$\lambda$6347 in
Fig.~\ref{SiII6347_fig}) and lines which clearly show a P\,Cyg profile
(e.g.\ Fe{\sc ii}$\lambda$6248 in Fig.~\ref{FeII6248_fig}) shows that
the variations are very similar, although the mean profiles of Si{\sc
  ii}$\lambda$6347 and Fe{\sc ii}$\lambda$6248 are very different. It
should be mentioned, however, that also the lines with pure absorption
profiles show clear variations from year to year in radial velocity
and line width.

\subsection{The resonance lines Na{\sc i}, Ca{\sc ii} and K{\sc i}}

The Na{\sc i}D lines show, apart from a strong interstellar component,
clearly also a stellar contribution: a weak emission component and
multiple variable absorption components. The variable absorption
components are evidence for accelerating expanding shells. The maximum
expansion velocity is typically around $-150$ km s$^{-1}$ and
reaches up to $-200$ km s$^{-1}$. Discrete components are frequently
observed in the velocity range from $-60$ to $-150$ km s$^{-1}$. 

\begin{figure}
  \resizebox{8.8cm}{!}{\includegraphics[angle=-90]{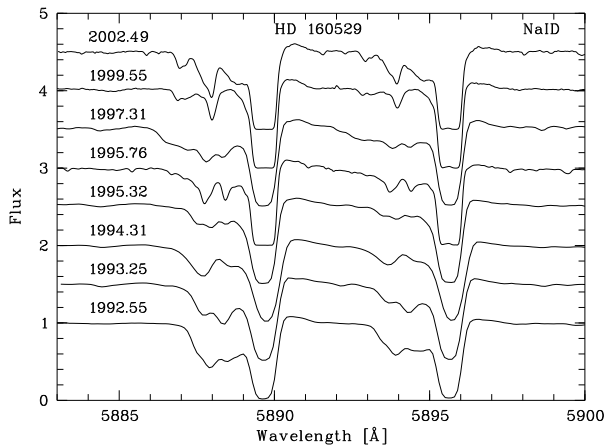}}
  \caption[]{Mean Na{\sc i} profiles in the years 1992, 1993, 1994,
    1995, 1997, 1999 and 2002. The strongest variations are seen at
    the blue edge of the absorption component. The spectra are labeled
    with {\bf date in years}.}
  \label{Na_all_fig}
\end{figure}

\begin{figure}
  \resizebox{8.8cm}{!}{\includegraphics[angle=0]{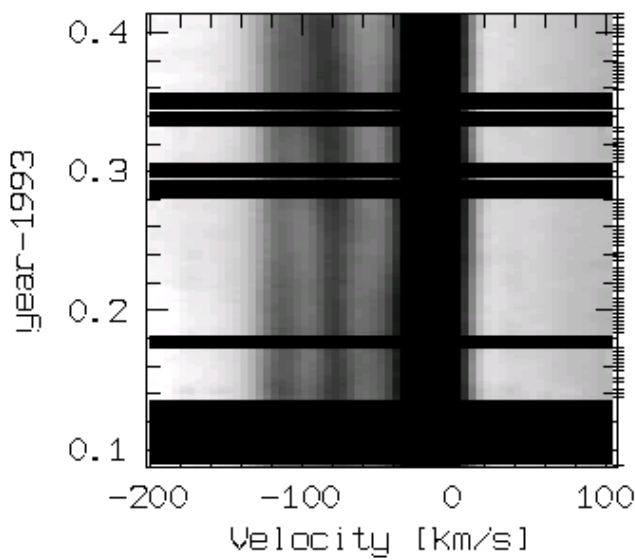}}
  \caption[]{Dynamic spectrum of Na{\sc i}$\lambda$5889 constructed 
    from the spectra obtained in 1993. Note the accelerating feature
    around $-75$ to $-90$ km s$^{-1}$. The acceleration of the
    feature is about 0.12 km s$^{-1}$ d$^{-1}$. }
  \label{NaI5889_fig}
\end{figure}

The changes in radial velocity of the components during one observing
season are relatively small {\bf (up to about about 15 km s$^{-1}$,
but typically below 10 km s$^{-1}$) The changes from year to year
are so large (cf.~Fig.~\ref{Na_all_fig}), that it is not clear how the
components are connected from year to year. Therefore, the
development of the velocities on longer time-scales cannot reliably be
determined from our observations. Similar components may be present at
lower velocities as well, but cannot be seen because of the strong
interstellar contribution, which saturates the absorption at
velocities smaller than about $-50$ km s$^{-1}$.} Acceleration of the
components in general is very small.  In 1992, no significant
acceleration was seen. In 1993 and 1995, an accelerating component was
seen, most clearly in 1993 (cf.~Fig.~\ref{NaI5889_fig}). Even in this
year, {\bf we observed an acceleration of about} 0.12 km s$^{-1}$
d$^{-1}$ at velocities from $-75$ to $-90$ km s$^{-1}$.  Another
component at higher velocities ($-115$ km s$^{-1}$) was stable in
radial velocity.  Both components first weakened and later
strengthened again.  In 1995, three components were observed ($-125$,
$-77$ and $-103$ to $-114$) and one was accelerating. In 1994, two
components were seen.  Both components were very stable in radial
velocity.  The component at large expansion velocity ($-120$ km
s$^{-1}$) was weakening, while the component at smaller velocity
($-75$ km s$^{-1}$) was getting stronger with time.

\begin{figure}
  \resizebox{8.8cm}{!}{\includegraphics[angle=-90]{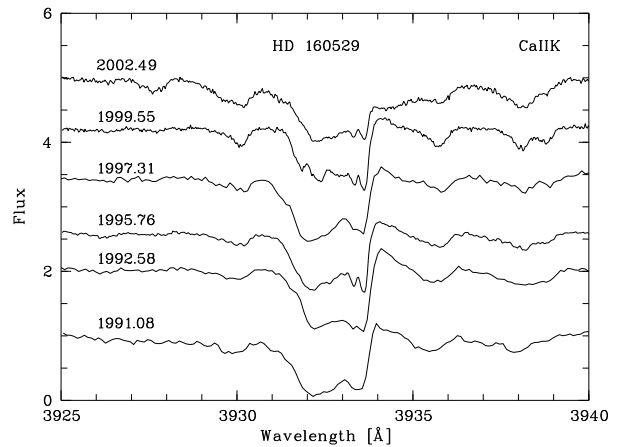}}
  \caption[]{Mean Ca{\sc ii} profiles in the years 1991, 1995, 1997,
    1999 and 2002. The strongest variations are seen at the blue edge
    of the absorption component. The spectra are labeled with 
    {\bf date in years}.}
  \label{Ca_all_fig}
\end{figure}

For Ca{\sc ii}H, K much less data is available. In particular, these
lines where not covered in the years 1992--1994, where we have
obtained the most extended time series. The yearly mean spectra of the
Ca{\sc ii}K line are shown in Fig.~\ref{Ca_all_fig}.  In cases where
both the Ca{\sc ii} and the Na{\sc i} lines are available, we find different
velocities in both groups of lines. The blue-shifted components are
much stronger in Ca{\sc ii} and therefore in most cases not as clearly
separated. For the un-shifted component, which is mainly of
interstellar origin, the Na{\sc i} lines are stronger than Ca{\sc ii}. In
Ca{\sc ii}, the interstellar contribution is split in two components
at $-26$ and $-4$ km s$^{-1}$. These components are only resolved in
the {\sc Ucles} and {\sc Feros} spectra.  This splitting is also
marginally visible in the Na{\sc i} lines.

Overall, the picture is not clear. {\bf There is no discernible
long-term trend in the velocity or acceleration of the components}.
From an expanding stellar wind ($R = 200$ R$_\odot$, $v_\infty = 200$
km s$^{-1}$, $\beta = 4$), the expected acceleration is about 1 km
s$^{-1}$ d$^{-1}$, i.e.\ significantly larger than what is
observed. This implies that the line-splitting is probably not due to
radially expanding shells moving with the stellar wind. {\bf This
behaviour is similar to what is observed for discrete absorption
components (DACs) in OB supergiants \citep[cf.~e.g.][]{fullerton:1997}. The origin of these features and their slow
acceleration is still not understood.}

The K{\sc i} line at 7699\,\AA\ is another interstellar absorption
line. This line has only been observed from 1995 to 2002.  Due to the
low abundance of K, the line is much weaker than Na{\sc i} or Ca{\sc
ii}. No blue-shifted components are seen.  The line has two absorption
components: a strong main component at $-2$ and a weaker component at
$-29$ km s$^{-1}$. The components are clearly resolved only in the
Feros spectra.  From the two Feros spectra it appears that the
fainter, blue component is slightly variable and therefore at least
partly of stellar origin.

\section{Stellar parameters}

\subsection{Distance}

{\bf The distance of HD\,160529} is important for the derivation of
stellar parameters. Unfortunately, the distance is very uncertain:
\citet{sterken:1991} estimate a distance of 2.5 kpc based on assigning
to HD\,160529 the same absolute visual magnitude as the LBV
\object{HD\,269662} (=R\,110) in the LMC ($M_\mathrm{V} = -
8.9$). R\,110 is spectroscopically very similar to HD\,160529 and also
has a comparable photometric amplitude.  In addition, an interstellar
extinction of $A_\mathrm{V}$ = 3.4 mag ($E_\mathrm{B-V} = 1.1$) is
assumed. The empirical relation between luminosity and amplitude of
LBVs \citep{wolf:1989} would give a lower bolometric luminosity of
about $M_\mathrm{Bol} = -8.3$, if we assume an amplitude of 0.5
mag. This would also yield a slightly smaller distance estimate {\bf
of about 1.9~kpc. If we assume that HD\,160529 has an absolute visual
brightness of $M_\mathrm{Bol} = -9.5$, similar to HD\,33579, the
brightest A hypergiant of the Large Magellanic Cloud, the distance
would be 3.5~kpc.}

A kinematic distance estimate is difficult for several reasons: the
systemic velocity is difficult to determine because of variable radial
velocities and line profile variations. In addition, the line-of-sight
to HD\,160529 is close to the galactic center and kinematic distances
therefore have large errors.  In the following, we will assume a
distance of 2.5 kpc but note that this distance has {\bf a rather
large uncertainty of about 30\%.}

\subsection{Temperature changes}

The temperature of HD\,160529 is another uncertain parameter. Line
strengths, derived e.g.\ from Si{\sc ii}, He{\sc i} and Fe{\sc ii} or
Ti{\sc ii} give contradictory results, when they are compared to
normal supergiants or analyzed with static model atmospheres. This is
certainly due to temperature stratification in the expanding
atmosphere and stellar wind contributions in most absorption lines.

The only metal which appears in different ionization stages is Mg.
The Mg{\sc i}/Mg{\sc ii} line ratio can be used to estimate the
spectral type.  The Mg{\sc i} lines $\lambda\lambda$3832, 3838, 5173,
5183 and the Mg{\sc ii} lines $\lambda\lambda$4481, 7877 were
used. The results are summarized in Table~\ref{Mg_tab}. For
comparison, the A0Ia star HD\,92207, the A2Ia star HD\,100262 and the
F3Ia star HD\,74180 were also measured. It should be noted that the
line profiles of the measured lines also change with time. Therefore,
the measured line ratios do not necessarily reflect changes in the
effective temperature {\bf only. It is likely, however, that most of
the variations of the line ratios reflect temperature changes.}

The near-IR also contains a large number of N{\sc i} lines, mainly in
the wavelength range around 7\,440 and 8\,700\,\AA\@. These lines are
very strong, significantly stronger than in any of the {\bf late
B/early-A supergiants} observed by \citet{przybilla:2001}. For the
spectra taken between 1995 and 1999, where the equivalent widths of
these lines are relatively stable, we obtain for the lines
$\lambda\lambda$7424, 7442, 7468 mean equivalent widths of 325, 417
and 533\,m\AA, respectively. In 2002 the lines are still stronger,
with equivalent widths of 423, 529 and 614\,m\AA, respectively.  For
HD\,100262 (A2Ia), we measure 87, 141 and 196\,m\AA, respectively. The
latter numbers are in very good agreement with the A2Ia star
HD\,111613 investigated by \citet{przybilla:2001}. The much stronger
lines for HD\,160529 may indicate an enhanced abundance of N, but
could also be due to non-LTE effects.

Lines from neutral oxygen are strong in HD\,160529. We measured an
equivalent width of between 2.15 and 2.40\,\AA\ for the near-IR triplet
$\lambda\lambda$7771-5. For A stars, the strength of this feature
increases with decreasing $\log g$ \citep{przybilla:2000}.

\begin{table}
  \begin{center}
  \caption[]{Mg{\sc i} and Mg{\sc ii} equivalent widths in
    m\AA\@. Note the strengthening of Mg{\sc i} in 1997 and later.
    The line ratio Mg{\sc i}/Mg{\sc ii} at most times gives a better
    match with HD\,100262 (A2Ia) than with HD\,92207 (A0Ia). In 2002,
    the equivalent widths are exceptionally high and indicate a much
    later spectral type.  The Mg{\sc i}$\lambda$3832 line is blended
    in HD\,74180.}
\begin{tabular}{l|rrrr|rr}
\multicolumn{1}{l}{} &
\multicolumn{4}{c}{Mg{\sc i}} &
\multicolumn{2}{c}{Mg{\sc ii}}\\
\hline
season & 3832  &  3838  & 5173  & 5183  &  4481  & 7877 \\
  \hline
1992 &  -    &   -    &  61   &  61   &   788  & - \\ 
1993 &  -    &   -    &  64   &  76   &   783  & - \\
1994 &  -    &   -    &  71   &  66   &   733  & - \\
1995 & 156   &   134  &  89   &  82   &   761  & 149 \\
1997 & 164   &   205  & 137   & 142   &   706  & 127 \\
1999 & 164   &   183  & 116   & 105   &   714  & 201 \\
2002 & 397   &   398  & 378   & 273   &   834  & 268 \\
\hline
HD\,92207 & 47 &  43   & 25    &  29   &   649  & 104 \\
HD\,100262 & 161 & 173   & 124  & 145  &   667  & 194 \\
HD\,74180  & -   & 450   & 553  & 491  &   766  & 220 \\
\hline
\end{tabular}
\label{Mg_tab}
\end{center}
\end{table}

In principle, the temperature can also be derived from the continuum
energy distribution. The optical energy distribution is not very
temperature sensitive, so the UV energy distribution is needed. In the
case of HD\,160529 this is difficult since the interstellar reddening
is high and quite uncertain.  However, low dispersion IUE spectra
obtained in 1992 and 1979, i.e.\ during the visual minimum and maximum
phase, respectively, show that the flux at about 2\,500\,\AA\ is {\em
lower\/} by about a factor 2 in visual maximum. This convincingly
demonstrates that the temperature was significantly lower at visual
maximum.

{\bf We tried to use synthetic spectra computed with the hydrostatic
model atmosphere code {\sc Tlusty} \citep{hubeny:1995,lanz:2001} to
determine the temperature of HD\,160529 and its changes. However, it
turned out that the line spectrum of HD\,160529 cannot be explained by
hydrostatic models. Models with solar abundance which fit the He{\sc
i}$\lambda$5876 line in minimum require about 11\,000~K, while fits of
metal lines such as, e.g., Mg{\sc ii}$\lambda$4481 would require about
9\,000~K. Increasing the He abundance reduces the discrepancy only
slightly. We interpret this result as a strong indication of
stratification effects in the expanding atmosphere. Non-LTE models
which include the atmospheric expansion and line blanketing are
required. While such models are now available for WR stars
\citep{hillier:1998}, they are so far not yet available for cooler
stars like HD\,160529.}

The changes of temperature and radius over the variability cycle can
be constrained by assuming that the bolometric magnitude of HD\,160529
is constant. The change of 0.5 magnitudes in the visual brightness
then reflects a change in the bolometric correction.  Assuming a
minimum temperature of 8\,000~K and a corresponding radius of 330
R$_\odot$, we derive a maximum temperature of 12\,000~K and a
corresponding radius of 150 R$_\odot$. The change of the temperature
is necessary for a corresponding change in bolometric correction. This
conclusion follows assuming a black-body distribution or if we use
empirical bolometric corrections \citep{schmidt-kaler:1982}. The large
change in radius is required by the constancy of the bolometric
luminosity. The temperatures correspond to spectral type of about B8
in visual minimum and A9 in visual maximum \citep{schmidt-kaler:1982},
in agreement with the spectroscopic evidence.

\subsection{Systemic velocity and rotational velocity}

The systemic velocity of HD\,160529 is difficult to determine since
all absorption lines are influenced to some extent by the stellar
wind, their radial velocities are variable and different for different
groups of lines. We therefore selected emission lines which appear
little disturbed by absorption to estimate the systemic velocity. The
Fe{\sc ii} lines of multiplets 40 and 46 appear well suited. From
these lines we derive a radial velocity of $-29 \pm 3$ km s$^{-1}$.
We also used the forbidden line N{\sc ii}$\lambda$6583 to estimate the
systemic velocity. This line is flat-topped and probably forms at
large distance from the star. The line center, as estimated from the
bi-sector of the left and right edge, is at $-8$ km s$^{-1}$, if we
use a rest wavelength of 6583.454\,\AA\ \citep{spyromilio:1995}.
Unfortunately, the line is faint and broad, so that the uncertainty of
this measurement is large.

The average velocity measured from He{\sc i}$\lambda$5876 is about
$-23$ km s$^{-1}$ (see below), which is relatively close to this
value. Other absorption lines such as the Cr{\sc ii} multiplet 30,
which appear as symmetric and narrow absorption lines in maximum, have
velocities of around $-20$ km s$^{-1}$.  The red O{\sc i} and N{\sc
i} lines give similar results: they vary between $-27$ and $-19$ km
s$^{-1}$ from 1995 to 2002. Because the radial velocity may be
variable also for symmetric lines, it is not clear whether this
velocity represents the true systemic velocity of HD\,160529.

Because of the variable line profiles, the rotational velocity $v \sin
i$ is also difficult to determine for HD\,160529. {\bf {\sc Feros}
spectra taken in 1999 and 2002 show particularly narrow absorption
lines. We used these spectra to estimate an upper limit for $v \sin
i$.  For this purpose, we fitted rotational profiles to some metal
lines, taking into account the instrumental resolution of the
spectrograph, but no other broadening mechanisms of the lines. In this
way, we derive a $v \sin i$ value of about 45 km s$^{-1}$. This has
to be considered as an upper limit, since stellar wind effects and
other broadening mechanisms are most likely important.}

\section{Modeling of H$\alpha$}

\label{secmodel}

Mass-loss rates derived from radio observations are only little
model-dependent. HD\,160529 is one of the few stellar objects
which are bright enough for current radio telescopes.
\citet{leitherer:1995} observed the star at 8.64 and 4.80\,GHz and
found that the measured flux is consistent with thermal emission from
an optically thick expanding wind at constant velocity.  The authors
adopt the spectroscopic distance of $d = 2.5$ kpc from
\citet{sterken:1991}, an electron temperature of
$T_\mathrm{e}$\,=\,5\,000\,K for the radio emitting region and a
terminal velocity of $v_\infty$\,=\,180\,km s$^{-1}$\ to derive a
mass-loss rate of
$\dot{M}$\,=\,$10^{-4.87\pm0.24}$\,$\mathrm{M}_\odot$yr$^{-1}$.

\begin{figure}
  \resizebox{8.8cm}{!}{\includegraphics[angle=-90,clip=t]{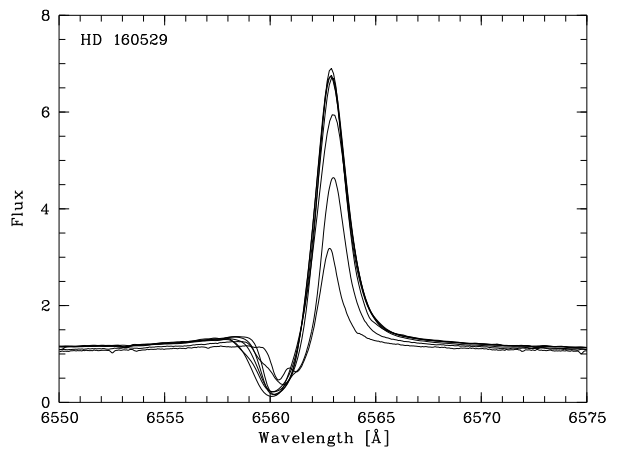}}
  \caption[]{Mean H$\alpha$ profile in the years 1992, 1993, 1994,
    1995, 1997, 1999 and 2002. In the years 1992--1995, the strongest
    variations are seen at the blue edge of the absorption component.
    After 1995, the emission decreased. The weakest emission is
    present in 2002, the second weakest in 1999. }
  \label{Ha_all_fig}
\end{figure}

From our observations (cf.~Fig.~\ref{Ha_all_fig}), we find that the
emission-line flux of HD\,160529 was slowly decreasing during our
observing period, i.e.\ with increasing visual brightness.  The
variations during each season are much smaller than this long-term
trend.

\begin{figure}
  \resizebox{8.8cm}{!}{\includegraphics[angle=0]{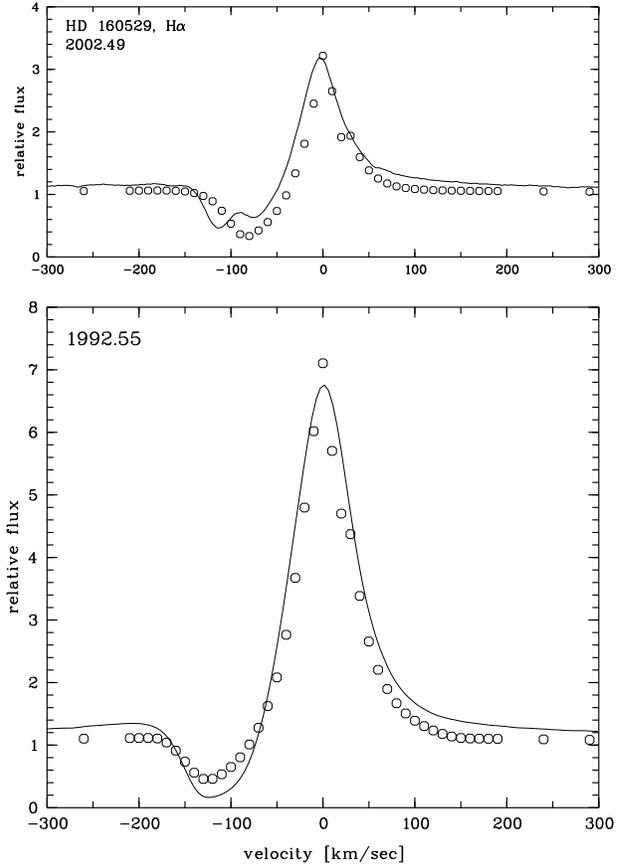}}
  \caption[]{Fit of the H$\alpha$ line in 1992 and 2002.}
  \label{Ha_sob}
\end{figure}

In order to quantitatively investigate the mass-loss rate of
HD\,160529 versus time, we modeled the mean H$\alpha$ line of the
years 1992 and 2002, which show the strongest and weakest H$\alpha$
emission, respectively.  We used the multi-level line-transfer code
described by \citet{bastian:1982} and \citet{stahl:1983}.  A terminal
velocity $v_\infty$ of 200 km s$^{-1}$ was used for both minimum and
maximum. {\bf Note that stellar-wind theory predicts a decrease of
$v_\infty$ with decreasing surface gravity, i.e.\ with increasing
visual brightness of LBVs. This is not easily observed in HD\,160529,
since there are no lines in the optical with a well-defined blue
absorption edge. The absorption edge in H$\alpha$ is variable, but
this can also be explained by changes in the optical depth of the
wind. A constant wind velocity cannot be excluded from our
observations. We used a $\beta$-type velocity law with $\beta$=4. A
velocity law with such a slow acceleration was found by
\citet{stahl:2001b} to fit the line profiles of \object{AG\,Car} in
outburst.} The stellar radius and effective temperature were estimated
from the photometry and the line spectrum of the given year.  Then we
only varied $\dot{M}$ and $v_\mathrm{ sys}$ until a satisfactory fit
was achieved. Solar abundances were assumed.  Possible clumping was
not taken into account.  The results are summarized in
Tab.~\ref{Ha_tab} and the fits are shown in Fig.~\ref{Ha_sob}.

We find that the mass-loss rate of HD\,160529 is almost independent of
temperature, i.e.\ visual brightness. Note that the derived mass-loss
rate increases slightly from minimum to maximum, although the
equivalent width decreases in this period. The reduced equivalent
width is mainly due to the radius increase and -- to a smaller extent
-- to the decrease in temperature. The radius increase causes (at
constant mass-loss rate) a density decrease which will decrease the
equivalent width of the Balmer lines. Therefore, if the radius
increase is smaller than derived from the assumption of constant
bolometric luminosity, the derived mass-loss rate could even slightly
decrease from minimum to maximum. {\bf The derived mass-loss rate is
in good agreement with the results of \citet{leitherer:1995}, which
are based on radio observations. We estimate that our mass-loss rates
are accurate within about a factor of two.  We therefore conclude that
within the errors the mass-loss rate did not change significantly
between minimum and maximum.}

{\bf A more advanced modeling of the stellar wind of HD\,160529 with
a code as e.g.\ described by \citet{hillier:1998} would certainly be
worthwhile. However, results of such models for cool stars such as
HD\,160529 are not yet published. }

\begin{table}
  \begin{center}
    \caption[]{Summary of the mass-loss rate determinations. $v_\mathrm{sys}
      = -10$ km s$^{-1}$ and $v_\infty = 200$ km s$^{-1}$ where used in
      both cases.}
      \begin{tabular}{lll}
	& min. & max \\
	\hline 
	$R/\mathrm{R}_\odot$ & 150 & 330\\
	$T_\mathrm{eff}/\mathrm{K}$ & 12\,000 & 8\,000 \\
	$M/\mathrm{M}_\odot\mathrm{yr}^{-1}$ &  $7\,10^{-6}$ &  $1\,10^{-5}$\\
      \end{tabular}
      \label{Ha_tab}
  \end{center}
\end{table}

\section{Discussion and conclusions}

For the interpretation of the short-term variability of HD\,160529,
the relevant time scales are the rotational time scale
$P_\mathrm{rot}$, the stellar-wind time scale $P_\mathrm{wind}$ and
the pulsation time scale $P_\mathrm{puls}$.

The time scales are defined by the following relations:

\begin{eqnarray}
2 \pi R_\ast / v_\mathrm{break}  < P_\mathrm{rot}  <  2 \pi R_\ast / v_\mathrm{rot} \label{eq:prot} \\
P_\mathrm{wind}       =   R_\ast/v_\infty \label{eq:pwind} \\ 
\log P_\mathrm{puls}  =   -0.275 M_\mathrm{Bol} - 
3.918 \log T_\mathrm{eff} + 14.543  \label{eq:ppuls}
\end{eqnarray}

In Equation~\ref{eq:prot}, the upper limit to $P_\mathrm{rot}$ is
estimated from $v \sin i$ assuming $\sin i = 1$. This value can also
be slightly larger because the derived $v \sin i$ value is an upper
limit.  {\bf Note that it is expected that the rotational velocity is
a function of phase. Because of the complex line profiles it could not
be determined in light minimum and is therefore set constant here.}
Because of the acceleration of the wind to its terminal velocity in
the extended stellar wind, the typical expansion time scale can be
significantly longer than given by Equation~\ref{eq:pwind}.  The
pulsation period for the fundamental radial mode given by
Equation~\ref{eq:ppuls} is the empirical fitting formula given by
\citet{lovy:1984}.

We used the following parameters from \citet{sterken:1991}:
$M_\mathrm{bol} = -8.9$, $M_\ast = 13\,\mathrm{M}_\odot$.  In addition,
we used $v_\infty$ = 200 km s$^{-1}$ and $v_\mathrm{rot}$ = 45 km
s$^{-1}$ for both maximum and minimum phase. For the stellar radius
and temperature we adopted 150 R$_\odot$/12\,000~K and 330
R$_\odot$/8\,000~K for minimum and maximum phase, respectively.  With
these parameters, we obtain the time scale estimates given in
Table~\ref{periodtab}.

\begin{table}
\caption[]{Summary of time scale estimates (in days) for the minimum
and maximum phase of HD\,160529.}
\begin{center}
\begin{tabular}{lll}
& min. (hot) & max. (cool) \\
\hline
$P_\mathrm{rot}$  & 59 \ldots 169 & 193 \ldots 371  \\
$P_\mathrm{wind}$ & $>$6          & $>$13           \\
$P_\mathrm{puls}$ & 10            & 49              \\
\end{tabular}
\label{periodtab}
\end{center}
\end{table}

Within the uncertainties, the time scales $P_\mathrm{wind}$ and
$P_{\rm puls}$ are compatible with the typical variation time scales
of 50--100 days, while the rotational time scale is significantly
longer.  However, {\bf since rotational modulation can also lead to
variations with an integer fraction of the rotational period, the time
scale of the variations cannot be used to distinguish the possible
mechanisms of the short-term variations in HD\,160529.}
The simultaneous variation of the optical brightness and the He{\sc
i}$\lambda$5876 line, however, is most easily explained by pulsations.
Pulsations have also been suggested as a cause for the photometric
micro-variations of LBVs \citep[cf.~e.g.][]{lamers:1998}.

The correlation of the equivalent width variations with visual
brightness on short time scales is in marked contrast with the
behaviour on long time scales. This strongly suggests that the
physical mechanism for the variations on short time scales is
different from the LBV-type variations.

It is rather surprising, that so far no strong dependence of the
observed time scale on the visual brightness -- and therefore the
radius -- has been detected. All plausible variability mechanisms
would predict such a dependence. It should be mentioned, however, that
\citet{sterken:1991} derived a period of 57 days in minimum while the
derived value in maximum was 101 days -- about a factor of two
longer. It is not clear, however, if the difference is due to a real
period change.

Most LBVs seem to increase their mass-loss rate in maximum.  However,
it is not clear that this is a general property of LBVs
\citep{leitherer:1997}. \citet{vink:2002} have shown that the mass-loss
rate of LBVs versus temperature carries important diagnostic
information. The mass-loss rate of \object{AG\,Car} versus phase has
been studied by \citet{stahl:2001b}. \citet{vink:2002} compared these
results with their predictions for line-driven stellar winds and found
a good agreement. They also predict that the mass-loss rate of LBVs is
a complicated function of temperature, surface gravity and
abundances. In particular, the mass-loss rate can even decrease with
decreasing temperature at temperatures below about 15\,000 K, in
particular at low masses. This predicted behaviour nicely fits the
observed behaviour of HD\,160529, which is an LBV with very low
temperature and mass.

Although the mass-loss rates of LBVs can be predicted with some
precision, the details of the mass-loss process are still not
understood. In particular the complicated line-profile variations and
the line splitting cannot be explained by radially expanding features.
Rotational modulation may be the cause for these variations.

\acknowledgements{We thank the European Southern Observatory (ESO) for
  the generous allocation of observing time and the staff at La Silla
  for the support in the installation of our {\sc Heros} instrument at
  the telescope.  We acknowledge the use of the SIMBAD database (CNRS
  data center, Strasbourg). This research has made use of NASA's
  Astrophysics Data System (ADS) Abstract Service. This work was
  supported by the Deutsche Forschungsgemeinschaft (Wo 296, 9-1).  TG
  would like to thank C.~Leitherer and acknowledges support from
  STScI-DDRF grants. This project was supported by the Belgian Federal
  Office for Scientific, Technical and Cultural Affairs (IUAP P5/36),
  the Belgian Fund for Scientific Research (FWO), and by the Danish
  Natural Science Research Council, partly through the center for
  Ground-Based Observational Astronomy. We express our gratitude to
  T.~Arentoft, A.~Bruch, H.W.~Duerbeck, H.~Melief, M.~Nolte and
  A.~Visser for contributing some photometric measurements in the
  framework of LTPV\@. We thank S.~Bagnuolo for taking some of the
  {\sc Feros} spectra. }

\bibliographystyle{aa}
\bibliography{aamnem99,mybib}
\end{document}